\documentclass[aps,prd,twocolumn,superscriptaddress,amsmath,amssymb,showpacs]{revtex4}

\usepackage{graphicx}
\usepackage{dcolumn}
\usepackage{bm}

\begin{document}

\title{Resonance and absorption spectra of the Schwarzschild black hole \\
for massive scalar perturbations: a complex angular momentum analysis}

\author{Yves D\'ecanini}
\email{decanini@univ-corse.fr} \affiliation{Equipe Physique
Th\'eorique, SPE, UMR 6134 du CNRS
et de l'Universit\'e de Corse,\\
Universit\'e de Corse, Facult\'e des Sciences, BP 52, F-20250 Corte,
France}

\author{Antoine Folacci}
\email{folacci@univ-corse.fr} \affiliation{Equipe Physique
Th\'eorique, SPE, UMR 6134 du CNRS
et de l'Universit\'e de Corse,\\
Universit\'e de Corse, Facult\'e des Sciences, BP 52, F-20250 Corte,
France} \affiliation{Centre de Physique Th\'eorique, UMR 6207 du
CNRS et des Universit\'es Aix-Marseille 1 et 2 et de l'Universit\'e
du Sud Toulon-Var, CNRS-Luminy Case 907, F-13288 Marseille, France}

\author{Bernard Raffaelli}
\email{raffaelli@univ-corse.fr} \affiliation{Equipe Physique
Th\'eorique, SPE, UMR 6134 du CNRS
et de l'Universit\'e de Corse,\\
Universit\'e de Corse, Facult\'e des Sciences, BP 52, F-20250 Corte,
France}

\date{\today}

\begin{abstract}

We reexamine some aspects of scattering by a Schwarzschild black
hole in the framework of complex angular momentum techniques. More
precisely, we consider, for massive scalar perturbations, the
high-energy behavior of the resonance spectrum and of the absorption
cross section by emphasizing analytically the role of the mass. This
is achieved (i) by deriving asymptotic expansions for the Regge
poles of the $S$-matrix and then for the associated weakly damped
quasinormal frequencies and (ii) by taking into account the analytic
structure of the greybody factors which allows us to extract by
resummation the physical information encoded in the absorption cross
section.

\end{abstract}

\pacs{04.70.-s, 04.50.Gh}

\maketitle

\section{Introduction}

Since the pioneering paper of Matzner \cite{Matzner1968} inspired by
a related but unpublished work of Hildreth \cite{Hildreth1964}, wave
scattering and absorption by black holes is a topic which has been
extensively studied due to (i) its mathematical interest because
scattering theory is a branch of mathematical physics (see, e.g.,
Ref.~\cite{New82}) which has found with black holes and curved
spacetimes a new and rich field of activities, but also and above
all due to (ii) its physical interest in connection with various
fundamental or experimental aspects of classical and quantum gravity
such as perturbation theory of black holes, gravitational wave
theory, quasinormal modes and resonant scattering theory, weak and
strong lensing, superradiance and instabilities, Hawking radiation,
information paradox, holography and CFT correspondence,
higher-dimensional field theories, analogue models of gravity... We
refer to the monograph of Futterman, Handler and Matzner
\cite{FuttermanEtAl1988} and to references therein for the
literature on wave scattering and absorption by black holes prior to
1988 and to
Refs.~\cite{AnninosETAL1992,Andersson1994a,Andersson1994b,DasGibbonsMathur1997,
MaldacenaStrominger1997,GlampedakisAndersson2001,CardosoETAL2004,
DolanDoranLasenby2006,CrispinoDolanOliveira2009,
DecaniniEspositoFareseFolacci2011} for a short and non-exhaustive
list of papers on this subject published since this date and which,
to our opinion, shed light on it from new and interesting points of
view.

Of course, in the context of scattering and absorption by black
holes, physicists have been mainly concerned with massless field
theories with spin $0$, $1/2$, $1$, $2$ which are considered to be
much more relevant, from a physical point of view, than massive
ones. However, regularly since the seventies, some interesting
articles dealing with massive field theories and considering, in
particular, the influence of the mass parameter on various aspects
of scattering in the frequency domain (scattering resonances, bound
states, cross sections, instabilities...) or in the time domain
(late-time tails, instabilities...) have been published (see, e.g.,
Refs.~\cite{CollinsDelbourgoWilliams1973,DeruelleRuffini1974,DeruelleRuffini1975,
DamourDeruelleRuffini1976,
Unruh1976,TernovETAL1978,ZourosEardley1979,Detweiler1980,Kofman1982,Gaina1989,Zaslavskii1990,SimoneWill1992})
and during the last decade an increasing number of papers dealing
with these same topics appeared (see, e.g.,
Refs.~\cite{KoyamaTomimatsu2001a,KoyamaTomimatsu2001b,Konoplya2002,XueWangSu2002,BurkoKhanna2004,
JungPark2004,OhashiSakagami2004,FuruhashiNambu2004,StrafussKhanna2005,KonoplyaZhidenko2005,
DoranLasenbyDolanHinder2005,LasenbyETAL2005,CardosoYoshida2005,
DolanDoranLasenby2006,Konoplya2006,KonoplyaZhidenkoMolina2007,
Dolan2007,GrainBarrau2008,Hod2011}).

In this paper, we shall consider the simple case of a massive scalar
field propagating on the Schwarzschild black hole and we shall
revisit two particular aspects of scattering by this gravitational
background in the framework of complex angular momentum (CAM)
techniques, i.e., in other words, by using the Regge pole machinery.
(For a review of CAM techniques in scattering theory, we refer more
particularly to the monograph of Newton \cite{New82} and for the use
of these techniques in the context of black hole physics, we refer
to
Refs.~\cite{Andersson1994a,Andersson1994b,DecaniniFJ_cam_bh,DecaniniFolacci2009,
DolanOttewill_2009,DecaniniFolacci2010a,DecaniniFolacciRaffaelli2010b,
DecaniniEspositoFareseFolacci2011,DolanOliveiraCrispino2011}). More
precisely, we intend to describe the high-energy behavior of the
scattering resonance spectrum and of the absorption cross section by
emphasizing the role of the mass parameter. It should be noted that
these two topics have been already considered by numerous authors
\cite{CollinsDelbourgoWilliams1973,Unruh1976,
Gaina1989,SimoneWill1992,JungPark2004,KonoplyaZhidenko2005,GrainBarrau2008}
but precise descriptions have been only obtained from purely
numerical analysis (see, however, Ref.~\cite{OhashiSakagami2004}
where an analytically solvable toy-model has been considered in
order to understand the mass-dependence of the scattering resonance
spectrum and a very recent paper by Hod \cite{Hod2011} where the
fundamental resonances of near-extremal Kerr black holes due to
massive scalar perturbations are derived analytically). The CAM
approach will permit us to go beyond numerical considerations and to
understand analytically in term of the mass parameter well-known
effects such as the migration of the complex quasinormal frequencies
and the behavior of the absorption cross section. Of course, it is
important to recall that Regge pole techniques are formally valid
for ``high" frequencies. As a consequence, they do not allow us to
describe in the usual CAM framework the existence of the bound state
spectrum
\cite{DeruelleRuffini1974,TernovETAL1978,Kofman1982,Zaslavskii1990,KonoplyaZhidenko2005,GrainBarrau2008}
which is an important aspect of the massive scalar field theory on
the Schwarzschild black hole appearing for rather low frequencies.

Our article is organized as follows. In Sec.~II, we shall first
construct high-frequency asymptotic expansions for the Regge poles
associated with the massive scalar field. We shall use two different
approaches, both emphasizing the role played by the black hole
photon sphere: (i) a powerful approach developed recently by Dolan
and Ottewill in Ref.~\cite{DolanOttewill_2009} which is based on a
novel ansatz for the Regge poles and the associated Regge modes and
(ii) a more traditional approach (see
Refs.~\cite{DecaniniFolacci2010a,DecaniniFolacciRaffaelli2010b} for
previous applications to massless theories) based on the WKB method
developed a long time ago by Schutz, Will and Iyer (see
Refs.~\cite{SchutzWill,Iyer1,Iyer2,WillGuinn1988,Konoplya2003}) to
study the resonant behavior of black holes. Then, from the Regge
poles, we shall obtain the weakly damped quasinormal frequencies of
the massive scalar field in term of the mass parameter. This will
allow us to show explicitly that when the mass of the scalar field
increases, the oscillation frequency of a quasinormal mode increases
while its damping decreases. In Sec.~III, we shall consider the
absorption problem for the massive scalar field. From Regge pole
techniques, we shall make a resummation of the absorption cross
section and then provide a simple formula describing very precisely,
at high energies, its behavior and emphasizing more particularly the
role of the mass parameter and of the black hole photon sphere. We
shall finally conclude this paper by briefly considering some
possible generalizations of the present work. Throughout this paper,
we shall use units such that $\hbar = c = G = 1$ and assume a
harmonic time dependence $\exp(-i\omega t)$ for the massive scalar
field.

\section{Regge poles for the massive scalar field and associated complex quasinormal frequencies}

\subsection{Generalities and notations}

In this subsection, we shall describe the first problem we intend to
solve by using the CAM machinery and we shall also fix the main
notations used in our article.

We first recall that the exterior of the Schwarzschild black hole of
mass $M$ is defined by the metric
\begin{equation}\label{Metric_Schwarzschild}
ds^2= -(1-2M/r)dt^2+ (1-2M/r)^{-1}dr^2+ r^2 d\sigma_2^2
\end{equation}
where $d\sigma_2^2=d\theta^2 + \sin^2 \theta d\varphi^2$ denotes the
metric on the unit $2$-sphere $S^2$ and with the Schwarzschild
coordinates $(t,r,\theta,\varphi)$ which satisfy $t \in ]-\infty,
+\infty[$, $r \in ]2M,+\infty[$, $\theta \in [0,\pi]$ and $\varphi
\in [0,2\pi]$.

In order to simplify discussions below and to interpret physically
some of our results, it is also necessary to point out various
aspects of scattering of massive (and massless) particles by the
Schwarzschild black hole linked more or less directly with the
existence of its photon sphere at $r=3M$ (see, e.g., Chap.~25 of
Ref.~\cite{MTW} or, for more precisions, Chap.~3 of
Ref.~\cite{Chandrasekhar_1983} and, for some of the notations we
shall use, Ref.~\cite{DolanDoranLasenby2006}). We consider a
particle with rest mass $\mu$ and energy $\omega > \mu$ and we
denotes by $p(\omega)=\sqrt{\omega^2-\mu^2}$ and by
$v(\omega)=p(\omega)/\omega$ the particle momentum and the particle
speed at large distances from the black hole. We note, in
particular, that
\begin{equation} \label{vitesse}
v(\omega)=\sqrt{1-\frac{\mu^2}{\omega^2}}.
\end{equation}
We recall that, associated with the parameter $\omega$ and for $\mu$
fixed, there exists a sphere located at $r=r_c(\omega)$ with
\begin{eqnarray} \label{crit_massive1}
&& r_c(\omega)=2M  \left( \frac{3+\left(1+8v^2(\omega)\right)^{1/2}}
{1+\left(1+8v^2(\omega)\right)^{1/2}} \right)
\end{eqnarray}
on which the massive particle can orbit the black hole on unstable
circular (timelike) geodesics. We have $r_c(\omega) \in ]3M,4M[$. We
also recall that the critical radius $r_c(\omega)$ defines a
critical impact parameter
\begin{eqnarray}\label{crit_massive2}
&& b_c(\omega)=\frac{M}{\sqrt{2} \, v^2(\omega)}\left[
8v^4(\omega)+20v^2(\omega)-1  \phantom{{\left(v^2(\omega)\right)^{3/2}}}  \right. \nonumber \\
&&  \qquad \qquad \qquad \qquad \qquad \left.
+\left(1+8v^2(\omega)\right)^{3/2} \right]^{1/2}.
\end{eqnarray}
The black hole captures any particle sent toward it with an impact
parameter $b<b_c(\omega)$ while particles with impact parameter
$b>b_c(\omega)$ are scattered. As a consequence, for particles with
rest mass $\mu$ and energy $\omega $, the geometrical cross section
of the Schwarzschild black hole is $\sigma_\mathrm{geo}(\omega)=\pi
b_c^2(\omega)$ or reads more explicitly
\begin{eqnarray} \label{crit_massive3}
& & \sigma_\mathrm{geo}(\omega)=\frac{\pi M^2}{2 \, v^4(\omega)}
\left[ 8v^4(\omega)+20v^2(\omega)-1  \phantom{{\left(v^2(\omega)\right)^{3/2}}}  \right. \nonumber \\
&&  \qquad \qquad \qquad \qquad \qquad \left.
+\left(1+8v^2(\omega)\right)^{3/2} \right].
\end{eqnarray}
It should be noted that $v(\omega)=1$ for $\mu = 0$ and from
Eqs.~(\ref{crit_massive1})-(\ref{crit_massive3}) we recover the
existence of the so-called black hole photon sphere located at
$r_c(\omega)=3M$ (the place on which the massless particles can
orbit the black hole on unstable circular null geodesics). Moreover,
in this case, the corresponding critical impact parameter is given
by $b_c(\omega)= 3\sqrt{3} M$ and, as a consequence, the geometrical
cross section $\sigma_\mathrm{geo}(\omega)=27\pi M^2$ of this black
hole for massless particles is also recovered. It should be also
noted that for $\mu \not= 0$ we have the asymptotic expansions
\begin{subequations}\label{AsymExp_rc_bc_sigma}
\begin{eqnarray}
& & r_c(\omega)= 3M \left[ 1+ \frac{\mu^2}{9\omega^2} +
\underset{\omega  \to
+\infty}{\mathcal{O}}\left(\frac{1}{\omega^4}\right) \right],
\\
 & & b_c(\omega)= 3\sqrt{3}M \left[1 +\frac{\mu^2}{3\omega^2} +
\underset{\omega  \to
+\infty}{\mathcal{O}}\left(\frac{1}{\omega^4}\right)   \right],  \\
& & \sigma_\mathrm{geo}(\omega)= 27\pi M^2 \left[1+\frac{
2\mu^2}{3\omega^2} + \underset{\omega  \to
+\infty}{\mathcal{O}}\left(\frac{1}{\omega^4}\right) \right].
\label{AsymExp_sigmaGEOM}
\end{eqnarray}
\end{subequations}
So, at high energy, a massive particle behaves as a massless one
and, in particular, orbits the black hole on unstable circular
geodesics very near the photon sphere.

From now on, we consider a massive scalar field $\Phi$ with mass
$\mu$ propagating on the exterior of the Schwarzschild black hole.
It satisfies the wave equation $(\Box-\mu^2) \Phi =0$ which reduces,
after separation of variables and the introduction of the radial
partial wave functions $\phi_{\omega, \ell}(r)$ with $\omega >0$ and
$\ell \in \mathbb{N}$, to the Regge-Wheeler equation
\begin{equation}\label{RW}
\frac{d^2 \phi_{\omega, \ell}}{dr_\ast^2} + \left[ \omega^2 -
V_{\ell}(r)\right] \phi_{\omega, \ell}=0.
\end{equation}
In Eq.~(\ref{RW}), $V_{\ell}(r)$ denotes the Regge-Wheeler potential
given by
\begin{equation}\label{pot_RW_Schw}
V_\ell(r) = \left(1-\frac{2M}{r} \right) \left[\mu^2+
\frac{(\ell+1/2)^2-1/4}{r^2} +\frac{2M}{r^3}\right]
\end{equation}
while $r_\ast$ is the so-called tortoise coordinate defined from the
radial Schwarzschild coordinate $r$ by $dr/dr_\ast=(1-2M/r)$. Here,
it is important to recall that the function $r_\ast=r_\ast(r)$
provides a bijection from $]2M,+\infty[$ to $]-\infty,+\infty[$.

In this paper, we shall focus on the {\it IN}-modes (see, e.g.,
Ref.~\cite{Boulware75} or Chap.~30 of Ref.~\cite{DeWitt03}) which
are the solutions of (\ref{RW}) with a purely ingoing behavior at
the event horizon $r=2M$, i.e., which satisfy
\begin{subequations} \label{bcRW}
\begin{equation}\label{bc1}
\phi_{\omega, \ell} (r) \underset{r_\ast \to -\infty}{\sim}
T_\ell(\omega) e^{-i\omega r_\ast}
\end{equation}
and which furthermore, at spatial infinity $r \to +\infty$, have an
asymptotic behavior of the form
\begin{eqnarray}\label{bc2}
& & \phi_{\omega, \ell}(r) \underset{r_\ast \to +\infty}{\sim}
 \left[ \frac{\omega}{p(\omega)}
\right]^{1/2}  \nonumber \\
& & \qquad \times \left(e^{-i[p(\omega)
r_\ast + (M\mu^2/p(\omega)) \ln(r/M)]}\right. \nonumber \\
& & \qquad \quad  \left. + R_\ell(\omega) e^{+i[p(\omega) r_\ast +
(M\mu^2/p(\omega)) \ln(r/M)]} \right).
\end{eqnarray}
\end{subequations}
In Eq.~(\ref{bc2}), $p(\omega)$ which now denotes the ``wave number"
is given by
\begin{equation}\label{p_de_om}
p(\omega)=\left( \omega^2 - \mu^2 \right)^{1/2}
\end{equation}
while $T_\ell(\omega)$ and $R_\ell(\omega)$ are transmission and
reflection coefficients linked by
\begin{equation}\label{NormCond}
|R_\ell(\omega)|^2+|T_\ell(\omega)|^2=1 \quad \forall \, \omega
>0 \,\, \mathrm{and} \,\, \forall \, \ell \in \mathbb{N}.
\end{equation}
It is interesting to note that this relation can be derived from the
properties of the Wronskian
\begin{eqnarray}\label{Wronskian}
& & W[\phi_{\omega, \ell}(r),{\overline {\phi_{\omega, \ell}}}(r)]
\equiv \phi_{\omega, \ell}(r) \left(\frac{d}{dr_\ast}{\overline
{\phi_{\omega, \ell}}}(r) \right)  \nonumber \\
& & \qquad\qquad\qquad     - \left(\frac{d}{dr_\ast}\phi_{\omega,
\ell}(r)\right){\overline {\phi_{\omega, \ell}}}(r).
\end{eqnarray}
Indeed, from (\ref{RW}) we can show that this Wronskian is a
constant and by evaluating it for $r_\ast \to -\infty$ and for
$r_\ast \to +\infty$ taking into account the boundary conditions
(\ref{bcRW}), we then obtain relation (\ref{NormCond}). It should be
also noted that the coefficient $\left[\omega /
p(\omega)\right]^{1/2}$ in Eq.~(\ref{bc2}) has been introduced in
order to simplify the form of this relation. It is moreover
important to remark that the {\it IN}-modes are naturally and
without any ambiguity defined by the boundary conditions
(\ref{bcRW}) for $\omega
> \mu$. However, for $0 < \omega < \mu$, the situation is a little
bit more complicated (see Chap.~30 of Ref.~\cite{DeWitt03} for more
precisions): indeed, it is first necessary to go into the complex
$\omega$ plane and to carefully take into account the various branch
cuts associated with the functions $p(\omega)$ (the two cuts
$]-\infty,-\mu[$ and $]+\mu,+\infty[$ along the real $\omega $ axis)
and $\omega ^{1/2}$ (e.g., a cut emanating from the origin and along
the negative imaginary $\omega$ axis) allowing us to deal with these
multivalued functions; then, by working on the Riemann sheet in
which $\mathrm{Im} \,p(\omega) \ge 0$ (the first Riemann sheet in
the following), we can define $p(\omega)$ for $0 < \omega < \mu$ and
we have in particular $p(\omega)$ which is a pure positive
imaginary; and, finally, the boundary conditions (\ref{bcRW}) can
now be used even for $0 < \omega < \mu$.

Moreover, it is worth recalling that the transmission coefficients
$T_\ell(\omega)$ permit us to construct the greybody factors (the
absorption probabilities by the Schwarzschild black hole for scalar
particles with energy $\omega$ and angular momentum $\ell$). They
are given by
\begin{equation}\label{Greybodyfactors}
\Gamma_\ell(\omega)= |T_\ell(\omega)|^2
\end{equation}
and, for the massive scalar field considered here, the black hole
absorption cross section can be expressed in terms of them in the
form
\begin{equation}\label{Sigma_abs}
\sigma_\mathrm{abs}(\omega)=\frac{\pi}{[p(\omega)]^2}
\sum_{\ell=0}^{+\infty} (2\ell + 1) \Gamma_\ell(\omega).
\end{equation}
Furthermore, even if we do not intend to analyze all the aspects of
scattering by this black hole and to consider in particular its
partial elastic cross sections, its scattering amplitude, its
differential cross section, etc. ..., it is interesting to recall
that all these concepts are built from the $S$-matrix which is
defined by its diagonal elements
\begin{equation}\label{S_matrix_def}
S_\ell(\omega)=(-1)^{\ell+1} R_\ell(\omega).
\end{equation}
Finally, it should be noted that the {\it IN}-modes alone do not
permit us to construct a basis of the solutions of the wave equation
but they are sufficient in order to describe the resonance and
absorption spectra of the Schwarzschild black hole. In other words,
here it is not necessary to consider the usual {\it UP}-modes or to
work with the alternative {\it OUT}- and {\it DOWN}-modes (see,
e.g., Chap.~4 of Ref.~\cite{Frolov-Novikov} or Chap.~30 of
Ref.~\cite{DeWitt03}).

Let us now consider the Regge-Wheeler potential $V_{\ell}(r)$ given
by (\ref{pot_RW_Schw}). From now on, we shall assume that the
condition
\begin{equation}\label{restriction_mu}
2M\mu  \in ]0,1/2[
\end{equation}
is satisfied. We shall therefore restrict our study to scalar field
mass $\mu$ rather weak. This assumption allows us to simplify our
study because it saves us from complicated discussions on the
behavior of $V_{\ell}(r)$ for different values of the angular
momentum $\ell$. Indeed, it is well-known
\cite{SimoneWill1992,GrainBarrau2008} that under
(\ref{restriction_mu}), $V_{\ell}(r)$ always presents three extrema
$\forall \ell \in \mathbb{N}$. Thanks to Tartaglia and Cardano, it
is furthermore easy to prove that these extrema
$r_{\mathrm{neg}}(\ell)$, $r_{\mathrm{max}}(\ell)$ and
$r_{\mathrm{min}}(\ell)$ satisfy $r_{\mathrm{neg}}(\ell)<0<8M/3 \le
r_{\mathrm{max}}(\ell) < 4M < r_{\mathrm{min}}(\ell)$ and are given
by
\begin{subequations} \label{extrema_potRW}
\begin{eqnarray}
&&r_{\mathrm{neg}}(\ell)=\frac{\ell(\ell+1)}{3M\mu^2}+2\sqrt{-\frac{\mathcal{P}(\ell)}{3}}
\cos{\left[\frac{\xi(\ell)}{3}+\frac{2\pi}{3}\right]},\qquad\\
&&r_{\mathrm{max}}(\ell)=\frac{\ell(\ell+1)}{3M\mu^2}+2\sqrt{-\frac{\mathcal{P}(\ell)}{3}}
\cos{\left[\frac{\xi(\ell)}{3}-\frac{2\pi}{3}\right]},\qquad \label{max_potRW}\\
&&r_{\mathrm{min}}(\ell)=\frac{\ell(\ell+1)}{3M\mu^2}+2\sqrt{-\frac{\mathcal{P}(\ell)}{3}}
\cos{\left[\frac{\xi(\ell)}{3}\right]},\qquad
\end{eqnarray}
\end{subequations}
where
\begin{subequations}
\begin{eqnarray}
&&{\mathcal P}(\ell)=\frac{-1}{3M^2\mu^4}\left[\ell^2(\ell+1)^2\right.\nonumber \\
&& \left. \qquad\qquad\quad -9M^2\mu^2\ell(\ell+1)+9M^2\mu^2\right]
\end{eqnarray}
and
\begin{eqnarray}
&&\xi(\ell)=\arccos\left[-\frac{{\mathcal
Q}(\ell)}{2}\sqrt{-\frac{27}{{\mathcal P}(\ell)^3}}\right]
\end{eqnarray}
with
\begin{eqnarray}
&&{\mathcal Q}(\ell)=\frac{-1}{27M^3\mu^6}\left[2\ell^3(\ell+1)^3-27M^2\mu^2\ell^2(\ell+1)^2\right.\nonumber \\
&& \left. \qquad \qquad  \quad
+27M^2\mu^2\ell(\ell+1)-216M^4\mu^4\right].
\end{eqnarray}
\end{subequations}
[For $\ell =0$, (\ref{extrema_potRW}) agrees with Eq.~(13) of
Ref.~\cite{GrainBarrau2008}.] Of course, only the extrema
$r_{\mathrm{max}}(\ell)$ and $r_{\mathrm{min}}(\ell)$ which lie in
the physical region $r> 2M$ govern the behavior of the {\it
IN}-modes defined above by the Regge-Wheeler equation (\ref{RW}) and
the boundary conditions (\ref{bcRW}) and, furthermore, it is very
important to keep in mind that $r_{\mathrm{max}}(\ell)$ corresponds
to the peak of a local potential barrier while
$r_{\mathrm{min}}(\ell)$ denotes the minimum of a local potential
well. Finally, it is interesting to note that
\begin{eqnarray}\label{DA du Max}
& & r_{\mathrm{max}}(\ell)=3M\left[ 1
-\frac{1-27M^2\mu^2}{9(\ell+1/2)^2} \right. \nonumber \\
& & \qquad \qquad \qquad \left. + \underset{\ell+1/2 \to
+\infty}{\cal O}\left(\frac{1}{(\ell+1/2)^4}\right) \right]
\end{eqnarray}
and therefore, for large angular momenta, the peak of the
Regge-Wheeler potential $V_\ell(r)$ lies very near the photon sphere
of the Schwarzschild black hole located at $r=3M$.

The process which has allowed us to define the {\it IN}-modes for $0
< \omega < \mu$ moreover permits us to extend for complex $\omega$
values the diagonal matrices $T$, $R$ and $S$. We recall that a pole
of $T_\ell(\omega)$ [let us note that it is also a pole of
$R_\ell(\omega)$ and $S_\ell(\omega)$] such that
$T_\ell(\omega)/R_\ell(\omega)$ remains regular and which lies in
the lower half plane of the first Riemann sheet associated with the
multivalued function $p(\omega)$ is a resonance of the scalar field.
We also recall that resonances are symmetrically distributed with
respect to the imaginary $\omega$ axis. We shall focus our attention
on those lying in the fourth quadrant of the considered Riemann
sheet. It is well-known that they can be separated into two families
(see, e.g., Chap.~30 of Ref.~\cite{DeWitt03} or
Refs.~\cite{DeruelleRuffini1974,TernovETAL1978,Kofman1982,Zaslavskii1990,
SimoneWill1992,KonoplyaZhidenko2005,GrainBarrau2008}) corresponding
respectively to complex frequencies $\omega$ satisfying $\mathrm{Re}
\, \omega < \mu$ and $\mathrm{Re} \, \omega
> \mu$ and therefore respectively associated with [see Eq.~(\ref{bcRW})]:

\qquad - a bound state spectrum (these modes are normalizable,
purely ingoing at the horizon and fall off exponentially at spatial
infinity),

\qquad - a quasinormal mode spectrum (these modes are not
normalizable, purely ingoing at the horizon and purely outgoing  at
spatial infinity and they oscillate at both boundaries).

From a physical point of view, the existence of the bound state
spectrum is directly related to the presence of the potential well
close to $r_{\mathrm{min}}(\ell)$ (see also
Refs.~\cite{DeruelleRuffini1974,TernovETAL1978,Kofman1982,Zaslavskii1990,GrainBarrau2008}
and Chap.~30 of Ref.~\cite{DeWitt03} for more precisions) while the
existence of the weakly damped quasinormal modes is due to the
presence of the potential barrier at $r_{\mathrm{max}}(\ell)$ or,
equivalently, can be semiclassically described in terms of ``surface
waves" lying close to the photon sphere at $r=3M$. In the remaining
of this section, we intend to discuss more precisely this last point
by using the CAM machinery. We are not be able to provide an
analogous description for the bound state spectrum.

To conclude this subsection, we shall briefly recall some aspects of
CAM techniques we shall extensively use in the following. We first
note that, for $\omega >0$, the matrices $T$, $R$ and $S$ previously
defined can be analytically extended into the CAM plane: we
transform the ordinary angular momentum $\ell$ into a complex number
$\lambda= \ell +1/2$ and we construct the analytic extensions
$T_{\lambda -1/2}(\omega)$, $R_{\lambda -1/2}(\omega)$ and
$S_{\lambda -1/2}(\omega)$ of $T_\ell(\omega)$, $R_\ell(\omega)$ and
$S_\ell(\omega)$ from (\ref{RW}), (\ref{pot_RW_Schw}), (\ref{bcRW})
and (\ref{S_matrix_def}). Let us also recall that the Regge poles
are defined as the poles of the $T$-matrix [or, equivalently, as the
poles of the matrices $R$ or $S$] for which
$T_{\lambda-1/2}(\omega)/R_{\lambda-1/2}(\omega)$ remains regular
and that they lye in the first and third quadrants of the CAM plane
symmetrically distributed with respect to its origin. The Regge
poles can be also considered as ``eigenvalues" associated with the
so-called Regge modes $\phi_{\omega, \lambda-1/2}(r)$ which satisfy
(\ref{RW}) and are purely ingoing at the horizon and purely outgoing
at spatial infinity [see Eq.~(\ref{bcRW})]. In
Ref.~\cite{DecaniniFolacci2010a} we showed that, for a massless
scalar field propagating on the Schwarzschild black hole, the
complex frequencies of the weakly damped quasinormal modes can be
obtain analytically from the so-called Regge trajectories, i.e.,
from the curves traced out in the CAM plane by the Regge poles as a
function of the frequency $\omega$. {\it Mutatis mutandis}, the
reasoning leading to these results (see for more precisions
Refs.~\cite{DecaniniFJ_cam_bh,DecaniniFolacci2010a} and Appendix A
of Ref.~\cite{DecaniniFolacciRaffaelli2010b}) can be repeated in the
context of a massive scalar field theory. Let us denote by
$\omega_{\ell n}=\omega^{(o)}_{\ell n}- i\Gamma_{\ell n}/2$ with
$\ell \in \mathbb{N}$ and $n\in
\mathbb{N}\setminus\lbrace{0\rbrace}$ the complex quasinormal
frequencies lying in the lower half plane of the first Riemann sheet
associated with the multivalued function $p(\omega)$ and by
$\lambda_n$ with $n \in \mathbb{N}\setminus\lbrace{0\rbrace}$ the
Regge poles lying in the first quadrant of the CAM plane. If we
describe the associated Regge trajectories by the functions
$\lambda_n=\lambda_n(\omega)$ with $\omega > \mu$, we have the
semiclassical relations
\begin{subequations} \label{sc12}
\begin{equation}\label{sc1}
\mathrm{Re}  \, \lambda_n \left(\omega^{(0)}_{\ell n} \right)= \ell
+ 1/2   \qquad \ell \in \mathbb{N},
\end{equation}
and
\begin{equation}\label{sc2} \frac{\Gamma _{\ell n}}{2}= \left.  \frac{\mathrm{Im} \, \lambda_n
(\omega )}{d/d\omega \, \ \mathrm{Re} \, \lambda_n (\omega )  }
\right|_{\omega =\omega^{(0)}_{\ell n}}.
\end{equation}
\end{subequations}

\subsection{Regge poles: The Dolan-Ottewill method}

In this subsection, we shall obtain high-frequency asymptotic
expansions for the Regge poles by using and extending to the massive
scalar field a new and powerful method introduced and developed by
Dolan and Ottewill in Ref.~\cite{DolanOttewill_2009}. It is
important to note (see below) that, in this method, the critical
parameters (\ref{crit_massive1}) and (\ref{crit_massive2})
associated with the massive scalar particle are the main ingredients
permitting us to construct the Regge modes and the Regge poles
associated with the field theory.

Extending the reasoning of Dolan and Ottewill to the massive scalar
field, we introduce the following ansatz to describe the Regge modes
$\phi_{\omega, \lambda_n(\omega)-1/2}(r)$ and the corresponding
Regge poles $\lambda_n (\omega)$:

\begin{widetext}
\begin{subequations} \label{DO-Ansatz1}
\begin{eqnarray} \label{DO-Ansatz1a}
& & \phi_{\omega, \lambda_n(\omega)-1/2}(r)=u_{\omega,
\lambda_n(\omega)-1/2}(r) \exp\left[i\omega v(\omega)
\int^{r_\ast}\left(1+\frac{2M
b_c(\omega)^2/r_c(\omega)^2}{r'}\right)^{\frac{1}{2}}\left(1-\frac{r_c(\omega)}{r'}\right)dr'_\ast\right]
\end{eqnarray}
with
\begin{eqnarray}\label{DO-Ansatz1b}
& & u_{\omega,
\lambda_n(\omega)-1/2}(r)=\left[\left(1-\frac{r_c(\omega)}{r}\right)^n
+\sum_{i=1}^{n}\sum_{j=1}^{\infty}b_{ij}^{(n)}(\omega)[\omega
v(\omega)]^{-j}\left(1-\frac{r_c(\omega)}{r}\right)^{n-i}\right]\exp\left(
\sum_{k=0}^{\infty} T_k^{(n)}(\omega,r)[\omega v(\omega)]^{-k}
\right) \nonumber \\
&&
\end{eqnarray}
\end{subequations}
and
\begin{equation} \label{DO-Ansatz2}
\lambda_{n}(\omega)=\lambda^{(n)}_{-1}(\omega v(\omega))\, [\omega
v(\omega)]+\lambda^{(n)}_{0}(\omega
v(\omega))+\frac{\lambda^{(n)}_{1}(\omega v(\omega))}{[\omega
v(\omega)]}+\frac{\lambda^{(n)}_{2}(\omega v(\omega))}{[\omega
v(\omega)]^2}+\frac{\lambda^{(n)}_{3}(\omega v(\omega))}{[\omega
v(\omega)]^3}+\frac{\lambda^{(n)}_{4}(\omega v(\omega))}{[\omega
v(\omega)]^4}+\ldots
\end{equation}
We invite the reader to compare our Eqs.~(\ref{DO-Ansatz1a}),
(\ref{DO-Ansatz1b}) and (\ref{DO-Ansatz2}) with Eqs.~(5), (39) and
(38) of Ref.~\cite{DolanOttewill_2009}. It should be noted more
particularly that, in order to extend the Dolan-Ottewill method for
the massive scalar field, we must consider as the natural asymptotic
parameter the ``momentum" $\omega v(\omega)=p(\omega)$ instead of
the energy $\omega$ and assume that $\omega > \mu$.

By inserting Eqs.~(\ref{DO-Ansatz1a}), (\ref{DO-Ansatz1b}) and
(\ref{DO-Ansatz2}) into (\ref{RW}) with (\ref{pot_RW_Schw}) where
$\ell \to \lambda_n(\omega)-1/2$ and after a tedious calculation, we
obtain
\begin{subequations}
\begin{eqnarray}
&&\lambda^{(n)}_{-1}(\omega v(\omega))=3\sqrt{3}M
\left[1+\frac{1}{3}\left(\frac{\mu^2}{[\omega
v(\omega)]^2}\right)-\frac{2}{27}\left(\frac{\mu^4}{[\omega
v(\omega)]^4}\right)+\underset{\mu/[\omega v(\omega)]
\to 0}{\cal O}\left(\frac{\mu^6}{[\omega v(\omega)]^6}\right)\right],\\
&&\lambda^{(n)}_{0}(\omega
v(\omega))=i\alpha(n)\left[1-\frac{1}{9}\left(\frac{\mu^2}{[\omega
v(\omega)]^2}\right)+\frac{1}{18}\left(\frac{\mu^4}{[\omega
v(\omega)]^4}\right)+\underset{\mu/[\omega v(\omega)]
\to 0}{\cal O}\left(\frac{\mu^6}{[\omega v(\omega)]^6}\right)\right],\\
&&\lambda^{(n)}_{1}(\omega
v(\omega))=\frac{3\sqrt{3}}{M}\left[\frac{60\alpha(n)^2-29}{11664}
+\frac{-372\alpha(n)^2+115}{104976}\left(\frac{\mu^2}{[\omega
v(\omega)]^2}\right)+\underset{\mu/[\omega v(\omega)]
\to 0}{\cal O}\left(\frac{\mu^4}{[\omega v(\omega)]^4}\right)\right],\\
&&\lambda^{(n)}_{2}(\omega
v(\omega))=i\frac{\alpha(n)}{M^2}\left[\frac{-1220\alpha(n)^2+1357}{419904}
+\frac{4444\alpha(n)^2-3707}{1259712}\left(\frac{\mu^2}{[\omega
v(\omega)]^2}\right)+\underset{\mu/[\omega v(\omega)] \to 0}{\cal
O}\left(\frac{\mu^4}{[\omega v(\omega)]^4}\right)\right],\\
&&\lambda^{(n)}_{3}(\omega
v(\omega))=\frac{3\sqrt{3}}{M^3}\left[\frac{-2357520\alpha(n)^4+4630008\alpha(n)^2-99373}{29386561536}
+\underset{\mu/[\omega v(\omega)]
\to 0}{\cal O}\left(\frac{\mu^2}{[\omega v(\omega)]^2}\right)\right],\\
&&\lambda^{(n)}_{4}(\omega
v(\omega))=i\frac{\alpha(n)}{M^4}\left[\frac{144920784\alpha(n)^4-439855800\alpha(n)^2+28395953}{2115832430592}
+\underset{\mu/[\omega v(\omega)] \to 0}{\cal
O}\left(\frac{\mu^2}{[\omega v(\omega)]^2}\right)\right],
\end{eqnarray}
\end{subequations}
and finally
\begin{eqnarray}\label{lambda_asymp}
&&\lambda_{n}(\omega )=3\sqrt{3}M\, \omega v(\omega)+i\alpha(n)
+\left[\frac{5}{36}\alpha(n)^2-\frac{29-3888M^2\mu^2}{432}\right]\left(\frac{1}{(3\sqrt{3}M\, \omega v(\omega))}\right)\nonumber\\
&& \qquad
+i\alpha(n)\left[-\frac{305}{3888}\alpha(n)^2+\frac{1357-46656M^2\mu^2}{15552}\right]
\left(\frac{1}{(3\sqrt{3}M \, \omega v(\omega))^2}\right)\nonumber\\
&& \qquad +\left[-\frac{49115}{839808}\alpha(n)^4+\frac{192917-4339008M^2\mu^2}{1679616}\alpha(n)^2 \right. \nonumber \\
&& \qquad\qquad  \left.
-\frac{99373-32192640M^2\mu^2+2176782336M^4\mu^4}{40310784}\right]\left(\frac{1}{(3\sqrt{3}M\, \omega v(\omega))^3}\right)\nonumber\\
&& \qquad+i\alpha(n)\left[\frac{3019183}{60466176}\alpha(n)^4
-\frac{18327325-311008896M^2\mu^2}{120932352}\alpha(n)^2 \right. \nonumber\\
&&\qquad\qquad   \left.
+\frac{28395953-6226336512M^2\mu^2+117546246144M^4\mu^4}{2902376448}\right]
\left(\frac{1}{(3\sqrt{3}M \, \omega v(\omega))^4}\right) \nonumber \\
& &\qquad +\underset{M\omega v(\omega) \to +\infty}{\cal
O}\left(\frac{1}{(3\sqrt{3}M\, \omega v(\omega))^5}\right).
\end{eqnarray}
In the previous equations, we have
\begin{equation}\label{alpha_def}
\alpha(n)=n-1/2 \quad \mathrm{for} \quad n\in
\mathbb{N}\setminus\lbrace{0\rbrace}.
\end{equation}

We can also convert the previous asymptotic expansion in $\omega
v(\omega)$ into an asymptotic expansion in $\omega$. From
(\ref{vitesse}) or (\ref{p_de_om}), we have
\begin{eqnarray}\label{PR_DO_def}
&& \lambda_n (\omega) =  3\sqrt{3}
M\omega+i\alpha(n)+\left[\frac{5}{36}\alpha(n)^2-\frac{29+1944M^2\mu^2}{432}\right]\left(\frac{1}{(3\sqrt{3}M\omega)}\right)\nonumber\\
&&+i\alpha(n)\left[-\frac{305}{3888}\alpha(n)^2+\frac{1357-46656M^2\mu^2}{15552}\right]\left(\frac{1}{(3\sqrt{3}M\omega)^2}\right)\nonumber\\
&&+\left[-\frac{49115}{839808}\alpha(n)^4+\frac{192917-1189728M^2\mu^2}{1679616}\alpha(n)^2
 -\frac{99373+4339008M^2\mu^2+952342272M^4\mu^4}{40310784}\right]\left(\frac{1}{(3\sqrt{3}M\omega)^3}\right)\nonumber\\
&&+i\alpha(n)\left[\frac{3019183}{60466176}\alpha(n)^4
-\frac{18327325-54867456M^2\mu^2}{120932352}\alpha(n)^2 \right. \nonumber \\
&& \qquad\qquad  \left.
+\frac{28395953+611380224M^2\mu^2-117546246144M^4\mu^4}{2902376448}\right]\left(\frac{1}{(3\sqrt{3}M\omega)^4}\right)
+ \underset{M\omega \to +\infty}{\cal O}
\left(\frac{1}{(3\sqrt{3}M\omega)^5}\right).
\end{eqnarray}

\end{widetext}
For $\mu=0$, formula (\ref{PR_DO_def}) is in agreement with Eq.~(38)
and Eqs.~(40)-(45) of Ref.~\cite{DolanOttewill_2009}.

\subsection{Regge poles: The WKB approach}

In this subsection, we shall derive again the high-frequency
asymptotic expansion (\ref{PR_DO_def}) for the Regge poles but, now,
we shall use a more traditional approach (see
Refs.~\cite{DecaniniFolacci2010a,DecaniniFolacciRaffaelli2010b} for
previous applications to massless theories) based on the WKB method
developed a long time ago by Schutz, Will and Iyer
\cite{SchutzWill,Iyer1,Iyer2,WillGuinn1988} (and extended to higher
orders by Konoplya \cite{Konoplya2003}) to study the resonant
behavior of black holes and to determine more particularly their
weakly damped quasinormal frequencies. We shall thus check formula
(\ref{PR_DO_def}) but we shall also see that the Dolan-Ottewill
method is, in the context of the Regge pole determination, a much
more powerful approach than the WKB one providing more quickly the
same results when we need to capture higher-order terms in
asymptotic expansions. Indeed, in order to derive (\ref{PR_DO_def})
we shall now start with a fifth-order WKB approximation for the
greybody factors (\ref{Greybodyfactors}) and, therefore, we shall
work with very heavy expressions (see below). Of course, if we only
need the leading order or the next-to-leading order of the
asymptotic expansion (\ref{PR_DO_def}), it seems to us that the WKB
approach remains more tractable.

Before to begin the technical part of our work, it is also
interesting to note that the Dolan-Ottewill method developed in the
previous subsection and the WKB approach we shall use here are based
on different physical concepts. Indeed, as already previously noted,
the Dolan-Ottewill ansatz is constructed from the critical
parameters (\ref{crit_massive1}) and (\ref{crit_massive2})
associated with the massive particle while the WKB calculation will
extensively use the maximum (\ref{max_potRW}) of the Regge-Wheeler
potential defining the scalar field theory. Of course, in both
approaches, for very high frequencies, it is the photon sphere at
$r_c(\omega)=3M$ and the corresponding impact parameter
$b_c(\omega)=3\sqrt{3}M$ which play the crucial role.

For $\ell \in \mathbb{N}$ and $\omega
>0$ with $\omega^2 $ near the peak $V_\ell(r_{\mathrm{max}}(\ell))$ of the Regge-Wheeler potential, we
can use, following Iyer, Will and Guinn
\cite{Iyer1,Iyer2,WillGuinn1988} and taking into account some
results displayed in Ref.~\cite{Konoplya2003}, a fifth-order WKB
approximation for the greybody factors (\ref{Greybodyfactors}).
After a tedious calculation, we obtain
\begin{equation}\label{Greybodyfactors_1} \Gamma_\ell(\omega)=
\frac{1}{1+\exp[2\mathcal{S}_\ell(\omega)]}
\end{equation}
with
\begin{widetext}
\begin{eqnarray}\label{RP_Phase_ordre5}
&&\mathcal{S}_\ell(\omega)=\pi k^{1/2}\left\{\frac{1}{2}z_0^2+\left(\frac{15}{64}b_3^2-\frac{3}{16}b_4\right)z_0^4+\left(\frac{1155}{2048}b_3^4-\frac{315}{256}b_3^2 b_4+\frac{35}{64}b_3 b_5+\frac{35}{128}b_4^2-\frac{5}{32}b_6\right)z_0^6\right. \nonumber \\
&& \left. \qquad \qquad \qquad +\left(\frac{255255}{131072}b_3^6-\frac{225225}{32768}b_3^4 b_4+\frac{15015}{4096}b_3^3 b_5+\frac{45045}{8192}b_3^2 b_4^2-\frac{3465}{2048}b_3^2 b_6-\frac{3465}{1024}b_3 b_4 b_5+\frac{315}{512}b_3 b_7 \right. \right. \nonumber \\
&& \left. \left. \qquad \qquad \qquad -\frac{1155}{2048}b_4^3+\frac{315}{512}b_4 b_6+\frac{315}{1024}b_5^2-\frac{35}{256}b_8\right)z_0^8+\left(\frac{66927861}{8388608}b_3^8-\frac{20369349}{524288}b_3^6 b_4+\frac{2909907}{131072}b_3^5 b_5 \right. \right. \nonumber \\
&& \left. \left. \qquad \qquad \qquad +\frac{14549535}{262144}b_3^4 b_4^2-\frac{765765}{65536}b_3^4 b_6-\frac{765765}{16384}b_3^3 b_4 b_5+\frac{45045}{8192}b_3^3 b_7-\frac{765765}{32768}b_3^2 b_4^3+\frac{135135}{8192}b_3^2 b_4 b_6 \right. \right. \nonumber \\
&& \left. \left. \qquad \qquad \qquad +\frac{135135}{16384}b_3^2 b_5^2-\frac{9009}{4096}b_3^2 b_8+\frac{135135}{8192}b_3 b_4^2 b_5-\frac{9009}{2048}b_3 b_4 b_7-\frac{9009}{2048}b_3 b_5 b_6+\frac{693}{1024}b_3 b_9+\frac{45045}{32768}b_4^4\right. \right. \nonumber \\
&& \left. \left. \qquad \qquad \qquad -\frac{9009}{4096}b_4^2 b_6-\frac{9009}{4096}b_4 b_5^2+\frac{693}{1024}b_4 b_8+\frac{693}{1024}b_5 b_7+\frac{693}{2048}b_6^2-\frac{63}{512}b_{10}\right)z_0^{10}\right\} \nonumber \\
&& \qquad +\pi k^{-1/2} \left\{\left(-\frac{7}{64}b_3^2+\frac{3}{16}b_4\right)+\left(-\frac{1365}{2048}b_3^4+\frac{525}{256}b_3^2 b_4-\frac{95}{64}b_3 b_5-\frac{85}{128}b_4^2+\frac{25}{32}b_6\right)z_0^2 \right. \nonumber \\
&& \left. \qquad \qquad \qquad +\left(-\frac{285285}{65536}b_3^6+\frac{315315}{16384}b_3^4 b_4-\frac{28875}{2048}b_3^3 b_5-\frac{79695}{4096}b_3^2 b_4^2+\frac{9765}{1024}b_3^2 b_6+\frac{8505}{512}b_3 b_4 b_5-\frac{1365}{256}b_3 b_7 \right. \right. \nonumber \\
&& \left. \left. \qquad \qquad \qquad +\frac{2625}{1024}b_4^3-\frac{1155}{256}b_4 b_6-\frac{1085}{512}b_5^2+\frac{245}{128}b_8\right)z_0^4+\left(-\frac{121246125}{4194304}b_3^8+\frac{43648605}{262144}b_3^6 b_4-\frac{7912905}{65536}b_3^5 b_5 \right. \right. \nonumber \\
&& \left. \left. \qquad \qquad \qquad -\frac{37011975}{131072}b_3^4 b_4^2+\frac{2777775}{32768}b_3^4 b_6+\frac{2477475}{8192}b_3^3 b_4 b_5-\frac{225225}{4096}b_3^3 b_7+\frac{2327325}{16384}b_3^2 b_4^3-\frac{585585}{4096}b_3^2 b_4 b_6 \right. \right. \nonumber \\
&& \left. \left. \qquad \qquad \qquad -\frac{555555}{8192}b_3^2 b_5^2+\frac{63525}{2048}b_3^2 b_8-\frac{525525}{4096}b_3 b_4^2 b_5+\frac{54285}{1024}b_3 b_4 b_7+\frac{49665}{1024}b_3 b_5 b_6-\frac{7035}{512}b_3 b_9-\frac{165165}{16384}b_4^4 \right. \right. \nonumber \\
&& \left. \left. \qquad \qquad \qquad +\frac{47355}{2048}b_4^2 b_6+\frac{45045}{2048}b_4 b_5^2-\frac{5985}{512}b_4 b_8-\frac{5355}{512}b_5 b_7-\frac{5145}{1024}b_6^2+\frac{945}{256}b_{10}\right)z_0^6\right\} \nonumber \\
&& \qquad +\pi k^{-3/2} \left\{\left(\frac{119119}{131072}b_3^6-\frac{153153}{32768}b_3^4 b_4+\frac{16107}{4096}b_3^3 b_5+\frac{47229}{8192}b_3^2 b_4^2-\frac{6405}{2048}b_3^2 b_6-\frac{6237}{1024}b_3 b_4 b_5+\frac{1155}{512}b_3 b_7\right. \right. \nonumber \\
&&\left. \left. \qquad \qquad \qquad -\frac{1995}{2048}b_4^3+\frac{1095}{512}b_4 b_6+\frac{1107}{1024}b_5^2-\frac{315}{256}b_8\right)+\left(\frac{156165009}{8388608}b_3^8-\frac{63864801}{524288}b_3^6 b_4+\frac{13216203}{131072}b_3^5 b_5\right. \right. \nonumber \\
&& \left. \left. \qquad \qquad \qquad +\frac{62777715}{262144}b_3^4 b_4^2-\frac{5354349}{65536}b_3^4 b_6-\frac{4945941}{16384}b_3^3 b_4 b_5+\frac{519057}{8192}b_3^3 b_7-\frac{4669665}{32768}b_3^2 b_4^3+\frac{1399167}{8192}b_3^2 b_4 b_6 \right. \right. \nonumber \\
&& \left. \left. \qquad \qquad \qquad +\frac{1368675}{16384}b_3^2 b_5^2-\frac{185661}{4096}b_3^2 b_8+\frac{1279047}{8192}b_3 b_4^2 b_5-\frac{161133}{2048}b_3 b_4 b_7-\frac{154917}{2048}b_3 b_5 b_6+\frac{28077}{1024}b_3 b_9 \right. \right. \nonumber \\
&& \left. \left. \qquad \qquad \qquad
+\frac{400785}{32768}b_4^4-\frac{143241}{4096}b_4^2
b_6-\frac{139293}{4096}b_4 b_5^2+\frac{23457}{1024}b_4 b_8
+\frac{22029}{1024}b_5
b_7+\frac{21777}{2048}b_6^2-\frac{5607}{512}b_{10}\right)z_0^2\right\}.
\end{eqnarray}
\end{widetext}
Here, we use the notations
\begin{eqnarray}\label{Greybodyfactors_3a}
& & z_0 \equiv z_0(\ell,\omega) = \sqrt{2 \, \frac{\omega^2 -
V_0(\ell)}{V^{(2)}_0(\ell)}}, \\
& & k \equiv k(\ell) = -\frac{1}{2} V^{(2)}_0(\ell),
\end{eqnarray}
and
\begin{eqnarray}\label{Greybodyfactors_3b}
&& b_p \equiv b_p(\ell) = \frac{2}{p!}
\frac{V^{(p)}_0(\ell)}{V^{(2)}_0(\ell)} \quad \mathrm{for} \quad
p>2,
\end{eqnarray}
with
\begin{equation} \label{V_0 et dersup1}
V_0(\ell) \equiv \left.
V_{\ell}(r_*)\right|_{r_*=r_*[r_{\mathrm{max}}(\ell)]} =\left.
V_{\ell}(r)\right|_{r=r_{\mathrm{max}}(\ell)}
\end{equation}
and
\begin{equation} \label{V_0 et dersup2}
V^{(p)}_0(\ell) \equiv \left. \frac{d^p}{{dr_*}^p }
V_{\ell}(r_*)\right|_{r_*=r_*[r_{\mathrm{max}}(\ell)]} \quad
\mathrm{for} \quad p \ge 2.
\end{equation}
It should be noted that, in Eq.~(\ref{RP_Phase_ordre5}), the term
$k^{1/2}z_0^2$ corresponds to the first-order WKB approximation.
Adding the terms $k^{1/2}z_0^4$ and $k^{-1/2}$ permits us to
constructed the second-order WKB approximation. Adding furthermore
the terms $k^{1/2}z_0^6$ and $k^{-1/2}z_0^2$ permits us to obtain
the third-order WKB approximation. Finally, adding the terms
$k^{1/2}z_0^8$, $k^{-1/2}z_0^4$ and $k^{-3/2}$ provides us with the
fourth-order WKB approximation. This fourth-order WKB approximation
can be found in Ref.~\cite{WillGuinn1988} (see Eq.~(12) of that
paper) and is confirmed by our own calculations, even if some
coefficients in Eqs.~(6) and (7) of Ref.~\cite{WillGuinn1988} are
wrong and if some misprints are present in Eq.~(4). In
Eq.~(\ref{RP_Phase_ordre5}) we have also taken into account the
terms $k^{1/2}z_0^{10}$, $k^{-1/2}z_0^6$ and $k^{-3/2}z_0^2$
corresponding to the fifth-order WKB approximation. They have been
obtained from Ref.~\cite{Konoplya2003}. It should be noted that we
have been able to construct also the sixth-order WKB approximation
of the greybody factors (\ref{Greybodyfactors_1}) by taking into
account the terms $k^{1/2}z_0^{12}$, $k^{-1/2}z_0^8$,
$k^{-3/2}z_0^4$ and $k^{-5/2}$. Because we do not need it here, we
do not display its very long expression but we can provide it upon
request.

Even if formula (\ref{RP_Phase_ordre5}) has been obtained for $\ell
\in \mathbb{N}$ and $\omega
>0$, it can be used in the complex frequency plane or, as we intend
to do here, in the complex angular momentum plane. We consider that
$\omega >0$ but we transform the angular momentum $\ell$ appearing
in the previous equations into the complex variable
$\lambda=\ell+1/2$ and we then consider the analytic extension
$\Gamma_{\lambda-1/2}(\omega)$ of the greybody factors
$\Gamma_\ell(\omega)$ defined by (\ref{Greybodyfactors}) and
(\ref{Greybodyfactors_1}) as well as the analytic extension
$\mathcal{S}_{\lambda-1/2}(\omega)$ of the ``phase"
$\mathcal{S}_\ell(\omega)$ given by (\ref{RP_Phase_ordre5}). We have
\begin{equation}\label{Greybodyfactors_Analytic} \Gamma_{\lambda-1/2}(\omega)=
\frac{1}{1+\exp[2\mathcal{S}_{\lambda-1/2}(\omega)]}
\end{equation}
and the (Regge) poles of the greybody factors are the solutions
$\lambda_n(\omega)$ of the equation
\begin{equation}\label{RP_phase}
\mathcal{S}_{\lambda-1/2}(\omega)= i(n-1/2) \pi \quad \mathrm{with}
\quad n \in \mathbb{N}\setminus\lbrace{0\rbrace}
\end{equation}
(here we only consider those of the poles lying in the first
quadrant of the CAM plane). By inserting (\ref{max_potRW}) into
(\ref{RP_Phase_ordre5})-(\ref{V_0 et dersup2}) and considering the
transformation $\ell \rightarrow \lambda-1/2$, we obtain from
(\ref{RP_phase})
\begin{widetext}
\begin{eqnarray}\label{EqWKB_ordre5}
& & (\omega M)^2 = \left(\frac{1}{27}\right) \lambda^2 -i
\left(\frac{2\alpha(n)}{27}\right) \lambda +\left[\left(\frac{29-276
\alpha(n)^2}{5832}\right) + \frac{1}{3}M^2\mu^2 \right]  \nonumber \\
& & + i  \left[\left(\frac{-1357 \alpha(n) +1220
\alpha(n)^3}{209952} \right)+ \frac{2 \alpha(n) }{9}
M^2\mu^2\right]\left(
\frac{1}{\lambda}\right)  \nonumber \\
& & + \left[\left(\frac{8545-736824 \alpha(n)^2-1193520 \alpha(n)
^4}{544195584} \right)+ \left(\frac{-7-60 \alpha(n)^2}{486}
\right)M^2\mu^2 + M^4\mu^4
\right]\left( \frac{1}{\lambda^2}\right)  \nonumber \\
&& + i  \left[\left(\frac{6835519 \alpha(n) -20410824 \alpha(n)
^3-25861584 \alpha(n)^5 }{39182082048}\right) + \left( \frac{-125
\alpha(n) -380 \alpha(n)^3}{7776} \right) M^2\mu^2
+   3\alpha(n) M^4\mu^4 \right] \left( \frac{1}{\lambda^3}\right) \nonumber \\
 & & + \underset{\lambda \to +\infty}{\cal O}
\left( \frac{1}{\lambda^4}\right).
\end{eqnarray}
\end{widetext}
We can then solve (\ref{EqWKB_ordre5}) perturbatively
to recover (\ref{PR_DO_def}).

\subsection{Regge poles: ``Exact" versus asymptotic results}

\begin{figure}
\includegraphics[height=6cm,width=8cm]{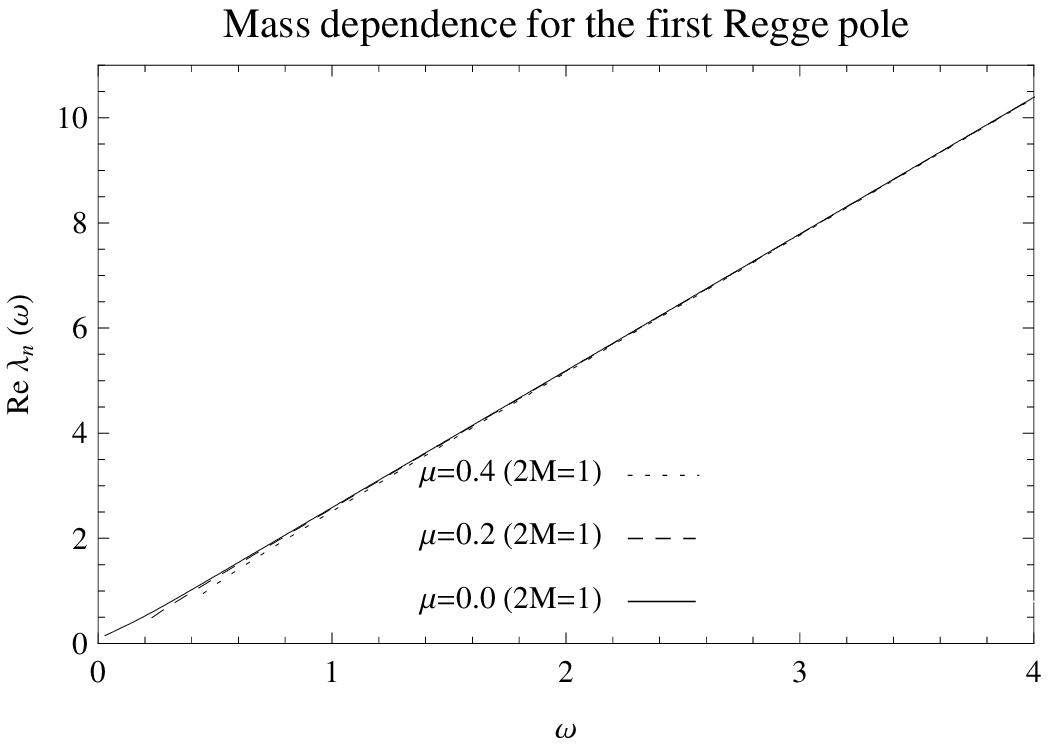}
\includegraphics[height=6cm,width=8cm]{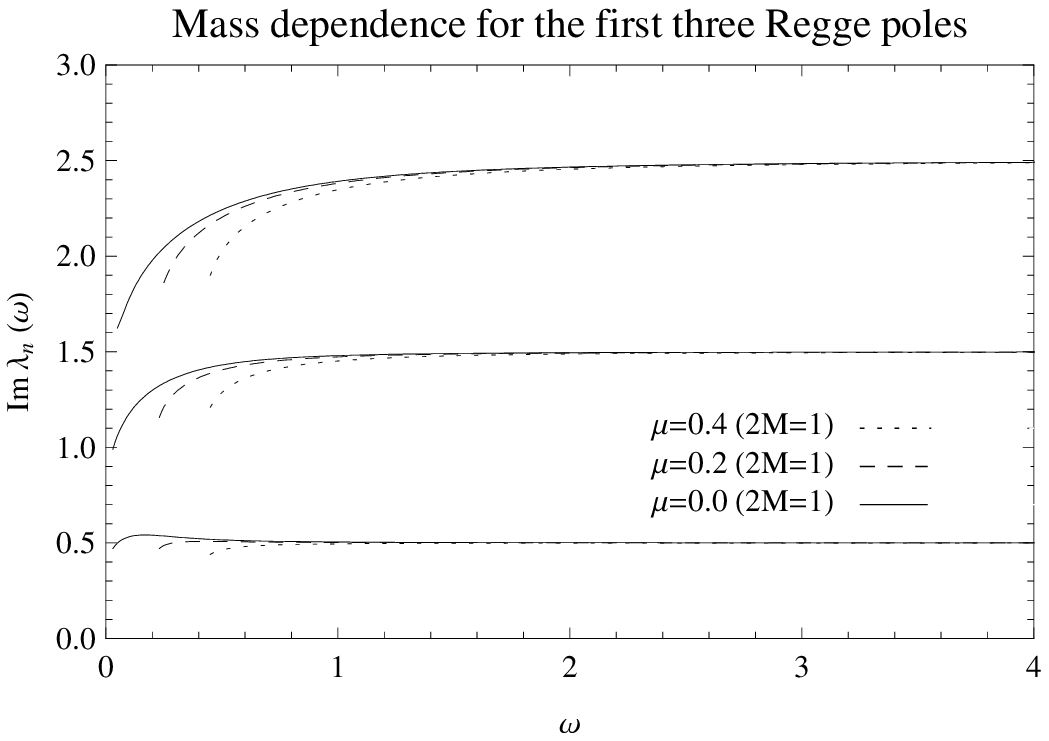}
\includegraphics[height=6cm,width=8cm]{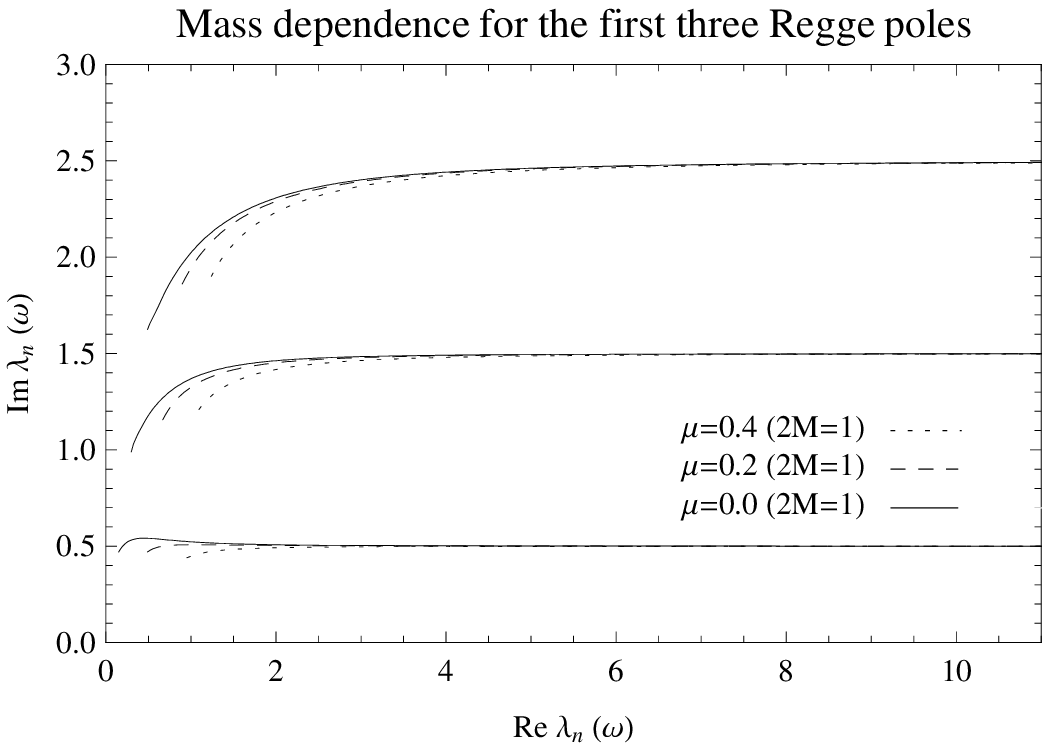}
 \caption{\label{fig:TRspin0}Regge trajectories for the massless and the massive scalar fields
 (small masses satisfying (\ref{restriction_mu})).
 The Regge poles $\lambda_n(\omega)$ ($n=1,2,3$) are followed
for $2M\omega=0.03 \to 4$. For more readability, the plot of
$\mathrm{Re} \, \lambda_n(\omega)$ is displayed for the only first
Regge pole.}
\end{figure}

It is now interesting to check the accuracy of the high-frequency
asymptotic expansions previously obtained. In
Ref.~\cite{DecaniniFJ_cam_bh}, in order to determine numerically,
for massless field theories, the Regge poles of the Schwarzschild
black hole, we adapted the powerful method developed by Leaver
\cite{LeaverI} to calculate the complex quasinormal frequencies of
the Schwarzschild and Kerr black holes (actually, we implemented a
slightly modified version of it due to Majumdar and Panchapakesan
and based on the Hill determinant \cite{mp}). But, here, we are
confronted with a massive scalar field and Leaver's method must be
modified. In Ref.~\cite{KonoplyaZhidenko2005}, Konoplya and Zhidenko
have shown how to achieve the corresponding modifications and, in
this paper, we have combined the Leaver-Konoplya-Zhidenko method
with the Hill determinant approach to deal numerically with the
Regge pole spectrum.

Figure \ref{fig:TRspin0} exhibits the exact Regge trajectories
numerically calculated and permits us to show clearly the role of
the mass parameter $\mu$. We also observe numerically that the Regge
poles are only defined for $\omega > \mu$.

We have also compared the exact Regge poles numerically obtained
with the results provided by the series (\ref{PR_DO_def}) or by the
series (\ref{lambda_asymp}). The accuracy is very impressive for
high frequencies (i.e. for $|2M\omega| \ge 2$) and the errors for
both series are of the same order. However, it should be noted that
they increase slightly with the order of the Regge pole and with the
magnitude of the mass parameter $\mu$. For rather low frequencies,
the series (\ref{lambda_asymp}) is the most accurate one.
Tables~\ref{tab:table1}-\ref{tab:table3} present a sample of Regge
pole positions calculated from this asymptotic formula. They exhibit
a good agreement for the first Regge pole even for low frequencies.
For the second Regge pole (and also for the third one not displayed
here), the accuracy remains correct for $|2M\omega| \ge 1$.

\begin{table}
\caption{\label{tab:table1} A sample of Regge pole values for the
scalar field: Exact versus asymptotic results for $\mu=0$ ($2M=1$).
The series (\ref{lambda_asymp}) is always truncated at the smallest
term.}
\begin{ruledtabular}
\begin{tabular}{cccccc}
&       &\quad Exact  \quad& \quad
Exact \quad&  Asymptotic  &  Asymptotic  \\
 $n $ & $\omega$ &$   \mathrm{Re} \, \lambda_n(\omega)  $
 &  $  \mathrm{Im} \, \lambda_n(\omega)   $
 & $  \mathrm{Re} \, \lambda_n(\omega)    $ &  $ \mathrm{Im} \, \lambda_n(\omega)   $  \\
\hline
1& 0.5 & 1.282821     & 0.515994  & 1.284398    & 0.515656  \\
 &  &    &   &  (-0.12 \%)  &  (0.07 \%) \\
 & 1.0 & 2.586845    & 0.504736  & 2.586891   & 0.504737 \\
  &  &    &   &    (-0.0018 \%)  &  (-0.0002 \%) \\
 & 2.0 & 5.190075    & 0.501236  & 5.190077   & 0.501236 \\
  &  &    &   &     (-0.000039 \%)  &  (0. \%) \\
2& 0.5 & 1.452818     & 1.430687  & 1.469628    & 1.379357 \\
  &  &    &   &    (-1.2 \%)  &  (3.6  \%) \\
 & 1.0 & 2.688355    & 1.479587  & 2.690232   & 1.477585 \\
  &  &    &   &     (-0.07 \%)  &  (0.14 \%) \\
 & 2.0 & 5.242994    & 1.494926  & 5.243088   & 1.494880 \\
  &  &    &   &     (-0.0018 \%)  &  (0.0031  \%) \\
\end{tabular}
\end{ruledtabular}
\end{table}
\begin{table}
\caption{\label{tab:table2} A sample of Regge pole values for the
scalar field: Exact versus asymptotic results for $\mu=0.2$
($2M=1$). The series (\ref{lambda_asymp}) is always truncated at the
smallest term.}
\begin{ruledtabular}
\begin{tabular}{cccccc}
&       &\quad Exact  \quad& \quad
Exact \quad&  Asymptotic  &  Asymptotic  \\
 $n $ & $\omega$ &$   \mathrm{Re} \, \lambda_n(\omega)  $
 &  $  \mathrm{Im} \, \lambda_n(\omega)   $
 & $  \mathrm{Re} \, \lambda_n(\omega)    $ &  $ \mathrm{Im} \, \lambda_n(\omega)   $  \\
\hline
1& 0.5 & 1.245875     & 0.506823  & 1.250055    & 0.504331  \\
 &  &    &   &  (-0.34 \%)  &  (0.49 \%) \\
 & 1.0 & 2.569229    & 0.502503  & 2.569344   & 0.502477 \\
  &  &    &   &    (-0.004 \%)  &  (0.005 \%) \\
 & 2.0 & 5.181378    & 0.500679  & 5.181381   & 0.500679 \\
  &  &    &   &     (-0.000066 \%)  &  (0. \%) \\
2& 0.5 & 1.411681     & 1.405188  & 1.4155982    & 1.345471 \\
  &  &    &   &    (-0.3 \%)  &  (4.2  \%) \\
 & 1.0 & 2.670066    & 1.473100  & 2.671532   & 1.471040 \\
  &  &    &   &     (-0.055 \%)  &  (0.14 \%) \\
 & 2.0 & 5.234201    & 1.493274  & 5.234282   & 1.493228 \\
  &  &    &   &     (-0.0015 \%)  &  (0.0030 \%) \\
\end{tabular}
\end{ruledtabular}
\end{table}
\begin{table}
\caption{\label{tab:table3} A sample of Regge pole values for the
scalar field: Exact versus asymptotic results for $\mu=0.4$
($2M=1$). The series (\ref{lambda_asymp}) is always truncated at the
smallest term.}
\begin{ruledtabular}
\begin{tabular}{cccccc}
&       &\quad Exact  \quad& \quad
Exact \quad&  Asymptotic  &  Asymptotic  \\
 $n $ & $\omega$ &$   \mathrm{Re} \, \lambda_n(\omega)  $
 &  $  \mathrm{Im} \, \lambda_n(\omega)   $
 & $  \mathrm{Re} \, \lambda_n(\omega)    $ &  $ \mathrm{Im} \, \lambda_n(\omega)   $  \\
\hline
1& 0.5 & 1.114489     & 0.462801  & 1.077877    & 0.429439  \\
 &  &    &   &  (3.3 \%)  &  (7.2  \%) \\
 & 1.0 & 2.514597    & 0.495198  & 2.514479   & 0.495068 \\
  &  &    &   &    (0.005 \%)  &  (0.026 \%) \\
 & 2.0 & 5.155079    & 0.498976  & 5.155077   & 0.498975 \\
  &  &    &   &     (0.00004 \%)  &  (0.00020 \%) \\
2& 0.5 & 1.258631     & 1.286677  & 1.55611   & 0.983336 \\
  &  &    &   &    (-24 \%)  &  (24  \%) \\
 & 1.0 & 2.613157    & 1.451939  & 2.611183   & 1.450803 \\
  &  &    &   &     (0.08 \%)  &  (0.08 \%) \\
 & 2.0 & 5.207607    & 1.488218  & 5.207596   & 1.488185 \\
  &  &    &   &     (0.00021 \%)  &  (0.00219 \%) \\
\end{tabular}
\end{ruledtabular}
\end{table}

\subsection{Effect of the mass on the complex quasinormal
frequencies}

Formulas (\ref{sc1}) and (\ref{sc2}) permit us to construct
analytical approximations for the resonance excitation frequencies
$\omega_{\ell n}^{(0)}$ and the damping $\Gamma_{\ell n}/2$ of the
quasinormal modes. By inserting into these two semiclassical
formulas the series (\ref{PR_DO_def}) truncated after the term in
$1/(3\sqrt{3}M\omega)^2$, we obtain
\begin{subequations}\label{QNMSch_sc}
\begin{eqnarray}
&&\omega_{\ell n}^{(0)}=
\frac{1}{3\sqrt{3}M}\left\{(\ell+1/2) \phantom{\left(\frac{1}{(\ell+1/2)}\right)} \right. \nonumber \\
& & \, \left.
+\left[-\frac{5}{36}\alpha(n)^2+\frac{29+1944M^2\mu^2}{432}\right]
\left(\frac{1}{(\ell+1/2)}\right) \right. \nonumber \\
& & \, \left. +\underset{\ell \to +\infty}{\cal
O}\left(\frac{1}{(\ell+1/2)^3}\right)\right\},\\
&&\frac{\Gamma_{\ell n}}{2}= \frac{\alpha(n)}{3\sqrt{3}M}\left\{1
\phantom{\left(\frac{1}{(\ell+1/2)}\right)} \right. \nonumber \\
& & \, \left.
+\left[\frac{235}{3888}\alpha(n)^2+\frac{313-116640M^2\mu^2}{15552}\right]
\left(\frac{1}{(\ell+1/2)^{2}}\right)   \right. \nonumber \\
& & \, \left. +\underset{\ell \to +\infty}{\cal
O}\left(\frac{1}{(\ell+1/2)^4}\right)\right\}.
\end{eqnarray}
\end{subequations}
It should be noted that, for $\mu=0$, formulas (\ref{QNMSch_sc}) are
in agreement with the results obtained in Ref.~\cite{Iyer2} (see
also Ref.~\cite{DecaniniFolacci2010a}).

From Eq.~(\ref{QNMSch_sc}), we can note that when the mass $\mu$ of
the scalar field increases, the oscillation frequency of a
quasinormal mode also increases while its damping decreases. This
well-known result observed numerically by various authors (see,
e.g., Refs.~\cite{SimoneWill1992,KonoplyaZhidenko2005}) is here
analytically described. It is moreover possible to provide a dual
explanation of this result by considering the physical
interpretation of the Regge poles in terms of ``surface waves"
located close to the photon sphere
\cite{Andersson1994b,DecaniniFJ_cam_bh,DecaniniFolacci2010a,DecaniniFolacciRaffaelli2010b},
even if we think that this description must taken with a grain of
salt. In this context, we recall that $\mathrm{Re}  \, \lambda_n
\left(\omega \right)$ represents the azimuthal propagation constant
of the $n$-th surface wave while $\mathrm{Im}  \, \lambda_n
\left(\omega \right)$ is its damping constant. From
(\ref{PR_DO_def}), it is then obvious that, for a fixed value of the
frequency $\omega$, all the azimuthal propagation constants and all
the damping constants decrease when the mass parameter $\mu$
increases. As a consequence, the resonance excitation frequencies
$\omega_{\ell n}^{(0)}$ which are the frequencies for which a
constructive interference due to the surface waves occurs [see also
Eq.~(\ref{sc1})] must necessarily increase with the mass parameter
$\mu$. Furthermore, because the attenuation of the $n$-th surface
wave decreases when the mass parameter $\mu$ increases, the energy
it radiates away during its repeated circumnavigations around the
black hole also decreases; it is then natural [see also
Eq.~(\ref{sc2})] to observe a similar behavior for the damping of
the associated quasinormal modes.

We can greatly improve the asymptotic expansions (\ref{QNMSch_sc})
by going beyond the semiclassical approach. Indeed, the complex
quasinormal frequencies are solutions of the equation
\begin{equation}\label{QNM_Beyond}
\lambda_n(\omega)= \ell +1/2 \quad \mathrm{with} \quad n \in
\mathbb{N}\setminus\lbrace{0\rbrace} \quad \mathrm{and} \quad \ell
\in \mathbb{N}.
\end{equation}
Such a result can be understood if we recall that a factor in $1/
\cos[\pi \lambda_n(\omega)]$ appears in all the residues series over
the Regge poles constructed from the CAM machinery (see, e.g.,
Eq.~(8) of Ref.~\cite{DecaniniFJ_cam_bh}). Equation
(\ref{QNM_Beyond}) can be solved perturbatively and we then obtain
\begin{widetext}
\begin{subequations}\label{QNMSch}
\begin{eqnarray}
&&\omega_{\ell n}^{(0)}=\frac{1}{3\sqrt{3}M}\left\{(\ell+1/2)
+\left[-\frac{5}{36}\alpha(n)^2+\frac{29+1944M^2\mu^2}{432}\right]\left(\frac{1}{(\ell+1/2)}\right)\right. \nonumber \\
&&\left.+\left[\frac{17795}{839808}\alpha(n)^4+\frac{18763-14346720M^2\mu^2}{1679616}\alpha(n)^2
-\frac{82283+20015424M^2\mu^2-136048896
M^4\mu^4}{40310784}\right]\left(\frac{1}{(\ell+1/2)^3}\right)\right.\nonumber
\\ &&\left.+\underset{\ell \to +\infty}{\cal
O}\left(\frac{1}{(\ell+1/2)^5}\right)\right\},
\end{eqnarray}
and
\begin{eqnarray}&&\frac{\Gamma_{\ell n}}{2}=\frac{\alpha(n)}{3\sqrt{3}M}
\left\{1+\left[\frac{235}{3888}\alpha(n)^2+\frac{313-116640M^2\mu^2}{15552}\right]\left(\frac{1}{(\ell+1/2)^{2}}\right)\right. \nonumber \\
&& \left. +\left[-\frac{234857}{60466176}\alpha(n)^4
-\frac{653125-953881920M^2\mu^2}{120932352}\alpha(n)^2 \right.
\right. \nonumber \\
& & \left. \left.  \qquad \qquad
-\frac{4832407-3269372544M^2\mu^2+29386561536M^4\mu^4}{2902376448}\right]
\left(\frac{1}{(\ell+1/2)^4}\right)+\underset{\ell \to +\infty}{\cal
O}\left( \frac{1}{(\ell+1/2)^6}\right)\right\}.
\end{eqnarray}
\end{subequations}
\end{widetext}
For $\mu=0$, formula (\ref{QNMSch}) is in agreement with Eq.~(7) and
Eqs.~(17)-(22) of Ref.~\cite{DolanOttewill_2009}.

\section{CAM description of the high-energy absorption cross section and role of the mass}

\subsection{Generalities}

In this section, we shall focus our attention on the absorption
cross section defined by (\ref{Sigma_abs}) (see also
Refs.~\cite{JungPark2004,GrainBarrau2008} for previous numerical
investigations). We shall more particularly provide a simple and
very accurate approximation of this series emphasizing the role of
the mass parameter $\mu$ and permitting us to describe qualitatively
and quantitatively its behavior for high frequencies/energies.

With this aim in mind, it is necessary to replace the sum over the
partial waves (\ref{Sigma_abs}) by the series over Regge poles
\begin{eqnarray}\label{sigma_abs_SC3}
& &  \sigma_\mathrm{abs}(\omega)=  \sigma_\mathrm{geo}(\omega)
\nonumber \\
& & \qquad  -\frac{4\pi^2}{[p(\omega)]^2}\,\mathrm{Re} \left(
\sum_{n=1}^{+\infty} \frac{ e^{i\pi[\lambda_n(\omega)-1/2]}\,
\lambda_n(\omega) \gamma_n(\omega)} {\sin[\pi
(\lambda_n(\omega)-1/2)]} \right) \nonumber \\
& & \qquad + \dots
\end{eqnarray}
In Eq.~(\ref{sigma_abs_SC3}), $\sigma_\mathrm{geo}(\omega)$ is the
geometrical cross section of the black hole given by
(\ref{crit_massive3}) and the $\lambda_n(\omega)$ with $n \in
\mathbb{N}\setminus\lbrace{0\rbrace}$ are those of the (Regge) poles
of the analytic extension $\Gamma_{\lambda-1/2}(\omega)$ of the
greybody factor $\Gamma_\ell(\omega)$ lying in the first quadrant of
the complex $\lambda$ plane while the $\gamma_n(\omega)$ are the
associated residues. We can obtain (\ref{sigma_abs_SC3}) from
(\ref{Sigma_abs}) by using the CAM machinery and by repeating for
$\omega
> \mu$,
{\it mutatis mutandis}, the main steps of Sec.~II of
Ref.~\cite{DecaniniEspositoFareseFolacci2011} where we considered
the absorption cross section for a massless scalar field. This is
possible due to the properties of the Regge poles and of the
analytic extension of the $T$-matrix mentioned in Sec.~II.A. In
Eq.~(\ref{sigma_abs_SC3}), the dots are associated with a background
integral along the imaginary $\lambda$ axis. We shall assume that it
can be neglected numerically and physically at high energies.

It is now possible to consider the high-energy behavior of the
absorption cross section (\ref{Sigma_abs}) by replacing into
(\ref{sigma_abs_SC3}) the functions $\lambda_n(\omega)$ and
$\gamma_n(\omega)$ by their high-frequency asymptotic expansions. We
have already at our disposal the expansion (\ref{PR_DO_def}) for the
Regge poles $\lambda_n(\omega)$. It remains to us to construct the
analogous expansions for the residues $\gamma_n(\omega)$.

\subsection{Regge-pole residues of the greybody factors}

It should be first noted that the Dolan-Ottewill method we have
found very efficient for the construction of the high-frequency
asymptotic expansion of the Regge poles cannot be naturally adapted
to the determination of the residue asymptotic expansions (see also
a remark below). However, it is possible to achieve such a job very
efficiently from the WKB approach of Sec.~II.C.

The Regge poles are the poles of the analytic extension
$\Gamma_{\lambda-1/2}(\omega)$ of the greybody factors given by
(\ref{Greybodyfactors_Analytic}) and the solutions of
Eq.~(\ref{RP_phase}). As a consequence, it is easy to prove that the
corresponding residues are given by
\begin{equation}\label{Res_phase}
\gamma_n(\omega)= \frac{-1/2}{\left. [d \,
\mathcal{S}_{\lambda-1/2}(\omega) / d\lambda
]\right|_{\lambda=\lambda_n(\omega)}}.
\end{equation}
Then, from (\ref{Res_phase}), (\ref{RP_Phase_ordre5}),
(\ref{max_potRW}) and (\ref{PR_DO_def}), we obtain the asymptotic
expansion (a fifth-order WKB approximation)
\begin{widetext}
\begin{eqnarray}\label{RESapproxWKB5}
&&\gamma_n(\omega) = -\frac{1}{2\pi} + i \alpha(n)\left[\frac{5}{36
\pi }\right]\left(\frac{1}{3\sqrt{3}M\omega}\right) +
\left[\frac{305 }{2592 \pi }\alpha^2(n)-\frac{1357-46656 M^2\mu ^2
}{31104 \pi }\right]\left(\frac{1}{(3\sqrt{3}M\omega)^2}\right)
\nonumber \\
& & + i\alpha(n) \left[\frac{49115}{419904 \pi
}\alpha^2(n)-\frac{192917-1189728 M^2\mu ^2 }{1679616 \pi }
\right]\left(\frac{1}{(3\sqrt{3}M\omega)^3}\right)
\nonumber \\
& & + \left[-\frac{15095915}{120932352 \pi }
\alpha^4(n)+\frac{18327325-54867456 M^2\mu ^2 }{80621568 \pi
}\alpha^2(n) \right. \nonumber
\\
& & \qquad \quad \left. -\frac {28395953 + 611380224 M^2\mu ^2  -
   117546246144 M^4\mu ^4 } {5804752896 \pi }
\right]\left(\frac{1}{(3\sqrt{3}M\omega)^4}\right)+
\underset{M\omega \to +\infty}{\cal
O}\left(\frac{1}{(3\sqrt{3}M\omega)^5}\right).
\end{eqnarray}
\end{widetext}

It is interesting to note that, in a very recent article
\cite{DolanOttewill2011}, Dolan and Ottewill have developed new
techniques based on their ansatz in order to construct analytically
the black hole excitation factors associated with the complex
quasinormal frequencies. Their techniques could be adapted to obtain
the two first terms of (\ref{RESapproxWKB5}) but, unfortunately,
this requires lot of work.

\subsection{CAM approximation for the absorption cross section}

Let us now insert (\ref{PR_DO_def}) and (\ref{RESapproxWKB5}) into
(\ref{sigma_abs_SC3}) and take into account
Eq.~(\ref{AsymExp_sigmaGEOM}). By noting that the contribution of
the Regge poles with $n >1$ is practically negligible and that, for
mathematical coherence, we only need the first three terms of
(\ref{PR_DO_def}) and the first two terms of (\ref{RESapproxWKB5}),
we obtain
\begin{widetext}
\begin{eqnarray}\label{Struct_Eik et fine_asymp}
&&\sigma_\mathrm{abs}(\omega) \approx 27\pi M^2 \left(1+\frac{2
\mu^2}{3\omega^2} -8\pi
e^{-\pi}\frac{\sin{[2\pi(3\sqrt{3}M)\omega]}}{2\pi(3\sqrt{3}M)\omega}
+16\pi e^{-2\pi}\frac{\sin{[4\pi(3\sqrt{3}M)\omega]}}{4\pi(3\sqrt{3}M)\omega}\right. \nonumber \\
&& \left. \qquad\qquad\qquad\qquad\qquad +\frac{4\pi^2
e^{-\pi}\left[-39+7\pi +972\pi
M^2\mu^2\right]}{27}\frac{\cos{[2\pi(3\sqrt{3}M)\omega]}}
{{\left[2\pi(3\sqrt{3}M)\omega\right]}^2}\right).
\end{eqnarray}
\end{widetext}

\begin{figure}
\includegraphics[height=6cm,width=8cm]{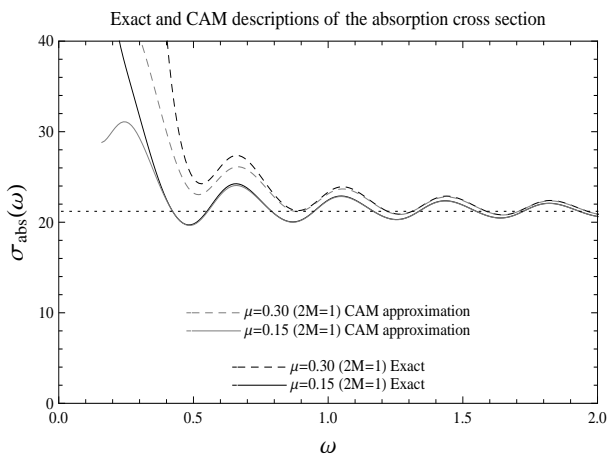}
 \caption{\label{fig:CAM absCS} The exact absorption cross section
 (\ref{Sigma_abs})
 and its CAM approximation (\ref{Struct_fine_asymp})
 are compared for $\mu= 0.15$ and $\mu=0.30$ ($2M=1$). The limiting constant value
 $27\pi M^2$ corresponding to the geometrical cross section of the photon sphere is
 also indicated.}
\end{figure}

\begin{figure}
\includegraphics[height=6cm,width=8cm]{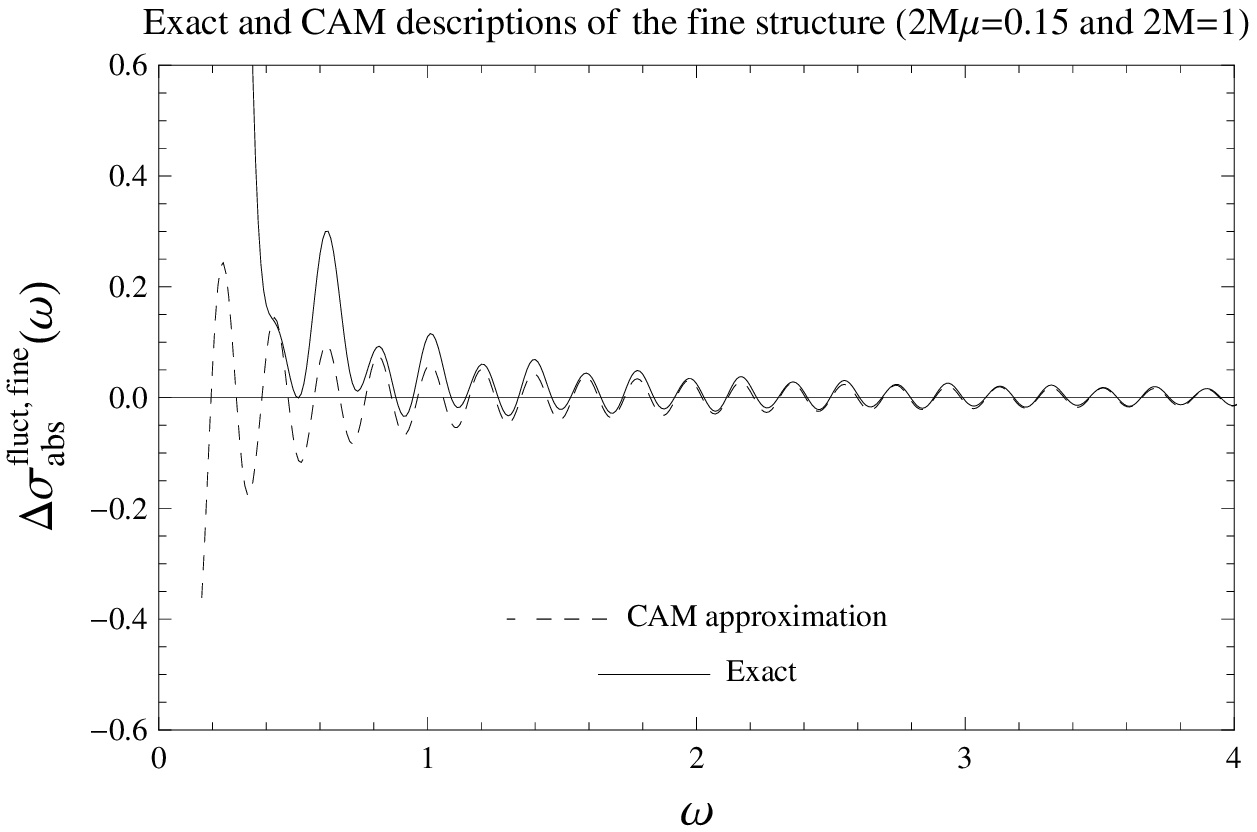}
\includegraphics[height=6cm,width=8cm]{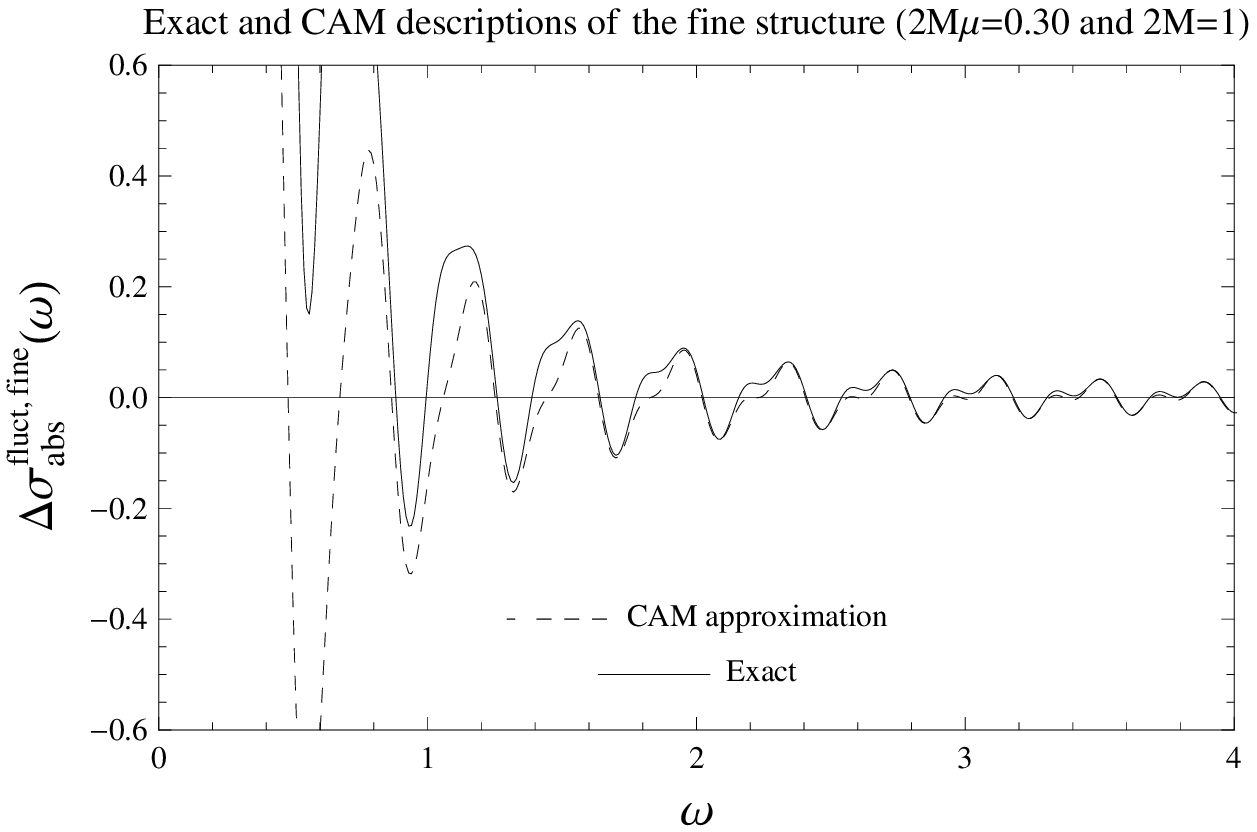}
 \caption{\label{fig:StructFine15et30} Fine structure for $\mu= 0.15$ and $\mu=0.30$ ($2M=1$). We compare the
 exact fine structure (\ref{Struct_fine_exact})
 with its CAM approximation (\ref{Struct_fine_asymp}).}
\end{figure}

For $\mu =0$, formula (\ref{Struct_Eik et fine_asymp}) reduces to
Eq.~(2.12) of Ref.~\cite{DecaniniFolacciRaffaelli2011a} and is the
superposition of an eikonal contribution (see also
Ref.~\cite{DecaniniEspositoFareseFolacci2011}) and a fine structure.
It is interesting to recall that the eikonal contribution is the sum
of the geometrical cross section of the black hole photon sphere and
a sinc function involving the geometrical characteristics (orbital
period and Lyapunov exponent) of the null unstable geodesics lying
on this photon sphere. It describes accurately the high-energy
behavior of the absorption cross section and, in particular, of its
regular and attenuated oscillations around a limiting value (see,
e.g., Ref.~\cite{Sanchez1978a} for a numerical analysis of this
behavior obtained a long time ago). The fine structure is a slightly
more complicated function which also involves the geometrical
characteristics of the null unstable geodesics lying on the photon
sphere and which, above all, permits us to capture small
fluctuations lying beyond the eikonal contribution. For $\mu \not=
0$, even if it is not so natural, an analogous description can be
provided: we can associated the first three terms of
(\ref{Struct_Eik et fine_asymp}), i.e.,
\begin{eqnarray}\label{Struct_eik}
& & \sigma^\mathrm{Eik}_\mathrm{abs}(\omega) \equiv 27\pi M^2
\left(1+\frac{2 \mu^2}{3\omega^2} \phantom{\frac{\sin{[(3\sqrt{3}M)]}}{(3\sqrt{3}M)}} \right. \nonumber \\
& & \qquad\qquad\qquad \left. -8\pi
e^{-\pi}\frac{\sin{[2\pi(3\sqrt{3}M)\omega]}}{2\pi(3\sqrt{3}M)\omega}
\right)
\end{eqnarray}
with the eikonal description and the two last ones, i.e.,
\begin{eqnarray}\label{Struct_fine_asymp}
&&27\pi M^2 \left[16\pi e^{-2\pi}\frac{\sin{[4\pi(3\sqrt{3}M)\omega]}}
{4\pi(3\sqrt{3}M)\omega}\right. \nonumber \\
&& \left.  +\frac{4\pi^2 e^{-\pi}\left[-39+7\pi +972\pi
M^2\mu^2\right]}{27}\frac{\cos{[2\pi(3\sqrt{3}M)\omega]}}
{{\left[2\pi(3\sqrt{3}M)\omega\right]}^2}\right]  \nonumber \\
& &
\end{eqnarray}
with the fine structure. We can observe that the mass parameter
$\mu$ appears in both contributions, greatly modifies the eikonal
part and slightly corrects the amplitude of one of the terms (the
smallest one for very high frequencies) of the fine structure.

We have tested numerically formula (\ref{Struct_Eik et fine_asymp}).
It describes very accurately the absorption cross section
(\ref{Sigma_abs}) as well as its oscillations around the geometrical
cross section and, in particular, it takes into account very
correctly the contributions of the mass parameter $\mu$. In
Fig.~\ref{fig:CAM absCS}, for two values of the mass parameter
$\mu$, we have displayed the exact absorption cross section
numerically obtained from (\ref{Sigma_abs}) by solving the problem
defined by (\ref{RW}), (\ref{pot_RW_Schw}) and (\ref{bcRW}) and we
have compared it with the result provided by (\ref{Struct_Eik et
fine_asymp}). The agreement is remarkable even for rather low
energies. We can however observe that, for low energies, the
accuracy of the CAM description decreases when the mass parameter
$\mu$ increases. Moreover, it should be noted that our numerical
absorption cross sections seem to be in agreement with those of
Ref.~\cite{GrainBarrau2008} but totally disagree with those
displayed in Ref.~\cite{JungPark2004}.

In Fig.~\ref{fig:StructFine15et30}, for the same values of the mass
parameter $\mu$, we have also tested the accuracy of the CAM
description (\ref{Struct_fine_asymp}) of the exact fine structure
defined by
\begin{eqnarray}\label{Struct_fine_exact}
& & \Delta \sigma_\mathrm{abs}^\mathrm{fluct, \, fine}(\omega)
\equiv \sigma_\mathrm{abs}(\omega) \nonumber \\
& & \quad -\left[27\pi M^2 \left(1+\frac{2 \mu^2}{3\omega^2} -8\pi
e^{-\pi}\frac{\sin{[2\pi(3\sqrt{3}M)\omega]}}{2\pi(3\sqrt{3}M)\omega}
\right) \right] \nonumber \\
& &
\end{eqnarray}
and numerically calculated. The agreement is remarkable for high
frequencies and remains robust even for rather low ones. It should
be noted that the behavior of the fine structure is very simple for
$2M\mu =0.15$ because, in this case, the coefficient of the second
term in Eq.~(\ref{Struct_fine_asymp}) can be neglected. In fact, it
vanishes for $2M\mu \approx 0.149265...$

\section{Conclusion and perspectives}

In the present paper, by working in the CAM plane, we have
emphasized explicitly the role of the mass parameter in the
resonance and absorption spectra of the Schwarzschild black hole for
massive scalar perturbations and simplified considerably their
description. Our work could be extended to the case of the massive
scalar field propagating on the Reissner-Nordstr\"om, Kerr and
Kerr-Newmann black holes as well as to the more interesting case,
from a physical point of view, of massive fermions \cite{Unruh1976}.
This last problem could even have nice applications in the context
of multimessenger high-energy astrophysics. With this aim in view,
the recent articles by Dolan, Doran, Lasenby and coworkers
\cite{DoranLasenbyDolanHinder2005,LasenbyETAL2005,DolanDoranLasenby2006}
would constitute a natural and solid starting point.

It is finally important to point out that, in this paper, the Regge
pole machinery has not permitted us to understand the existence of
the bound state spectrum
\cite{DeruelleRuffini1974,TernovETAL1978,Kofman1982,Zaslavskii1990,
SimoneWill1992,KonoplyaZhidenko2005,GrainBarrau2008}
associated with the massive scalar field theory defined on the
Schwarzschild black hole. This is really a pity and we consider that
it is one of the crucial problems to be tackled in the framework of
CAM techniques. Indeed, if we could solve such a problem, we could
maybe provide a dual simple explanation of the realization by
massive fields of the so-called ``black hole bomb" scenario
\cite{PressTeukolsky1972} or, in other words, of the instabilities
induced by the bound state spectrum in the presence of rotation and
due to the superradiance phenomenon (see, e.g.,
Refs.~\cite{DeruelleRuffini1975,DamourDeruelleRuffini1976,ZourosEardley1979,
Detweiler1980,FuruhashiNambu2004,
StrafussKhanna2005,CardosoYoshida2005,Dolan2007}).

\begin{acknowledgments}

We are grateful to Alexei Starobinsky for information concerning the
Russian literature and for having pointed to our attention
Refs.~\cite{TernovETAL1978} and \cite{Kofman1982}. We also thank Sam
Dolan for providing us with recent references on black hole
instabilities.

\end{acknowledgments}


\bibliography{RP_Schw_Mass}

\end{document}